\documentclass[
superscriptaddress,
reprint,
amsmath,amssymb,
aps, physrev,
floatfix,
]{revtex4-2}
\usepackage[english]{babel}
\usepackage[
    colorlinks=true,
    frenchlinks=false,
    linkcolor=blue,
    anchorcolor=blue,
    citecolor=blue,
    filecolor=blue,
    urlcolor=blue,
    bookmarks=true,
    bookmarksopen=true,
    bookmarksnumbered=true,
    bookmarksopenlevel=1,
    plainpages=false,
    pdfpagelabels=true,
    draft=false]{hyperref} 

\usepackage{graphicx} 
\usepackage[separate-uncertainty=true, multi-part-units=single]{siunitx} 
\usepackage{color} 
\usepackage{braket}
\usepackage{bbold}
\DeclareSIUnit\dBm{dBm}
\DeclareSIUnit\bar{bar}
\usepackage[mathlines]{lineno} 
\begin{document}

\title{Quantum teleportation over thermal microwave network}

\author{W.~K.~Yam}
\email{WunKwan.Yam@wmi.badw.de}
\affiliation{Walther-Mei{\ss}ner-Institut, Bayerische Akademie der Wissenschaften, 85748 Garching, Germany}
\affiliation{School of Natural Sciences, Technical University of Munich, 85748 Garching, Germany}

\author{S.~Gandorfer}
\affiliation{Walther-Mei{\ss}ner-Institut, Bayerische Akademie der Wissenschaften, 85748 Garching, Germany}
\affiliation{School of Natural Sciences, Technical University of Munich, 85748 Garching, Germany}

\author{F.~Fesquet}
\affiliation{Walther-Mei{\ss}ner-Institut, Bayerische Akademie der Wissenschaften, 85748 Garching, Germany}
\affiliation{School of Natural Sciences, Technical University of Munich, 85748 Garching, Germany}

\author{M.~Handschuh}
\affiliation{Walther-Mei{\ss}ner-Institut, Bayerische Akademie der Wissenschaften, 85748 Garching, Germany}
\affiliation{School of Natural Sciences, Technical University of Munich, 85748 Garching, Germany}

\author{K.~E.~Honasoge}
\affiliation{Walther-Mei{\ss}ner-Institut, Bayerische Akademie der Wissenschaften, 85748 Garching, Germany}
\affiliation{School of Natural Sciences, Technical University of Munich, 85748 Garching, Germany}

\author{A.~Marx}
\affiliation{Walther-Mei{\ss}ner-Institut, Bayerische Akademie der Wissenschaften, 85748 Garching, Germany}

\author{R.~Gross}
\affiliation{Walther-Mei{\ss}ner-Institut, Bayerische Akademie der Wissenschaften, 85748 Garching, Germany}
\affiliation{School of Natural Sciences, Technical University of Munich, 85748 Garching, Germany}
\affiliation{Munich Center for Quantum Science and Technology (MCQST), 80799 Munich, Germany}

\author{K.~G.~Fedorov}
\email{Kirill.Fedorov@wmi.badw.de}
\affiliation{Walther-Mei{\ss}ner-Institut, Bayerische Akademie der Wissenschaften, 85748 Garching, Germany}
\affiliation{School of Natural Sciences, Technical University of Munich, 85748 Garching, Germany}
\affiliation{Munich Center for Quantum Science and Technology (MCQST), 80799 Munich, Germany}

\begin{abstract}
Quantum communication in the microwave regime is set to play an important role in distributed quantum computing and hybrid quantum networks. However, typical superconducting quantum circuits require millikelvin temperatures for operation, which poses a significant challenge for large-scale microwave quantum networks. Here, we present a solution to this challenge by demonstrating the successful quantum teleportation of microwave coherent states between two spatially-separated dilution refrigerators over a thermal microwave channel in the temperature range up to $\SI{4}{\kelvin}$. We distribute two-mode squeezed states over this noisy channel and employ the resulting quantum entanglement for quantum teleportation of coherent states with fidelities of $\SI{72.3 \pm 0.5}{\percent}$ at $\SI{1}{\kelvin}$ and $\SI{59.9 \pm 2.5}{\percent}$ at \SI{4}{\kelvin}, exceeding the no-cloning and classical communication thresholds, respectively. We successfully model the teleportation protocol using a Gaussian operator formalism that includes losses and noise. Our analysis shows that the teleportation infidelity mainly stems from a parasitic heating of the cold quantum nodes due to the hot network connection. These results demonstrate the experimental feasibility of distributed superconducting architectures and motivate further investigations of noisy quantum networks in various frequency regimes.
\end{abstract}

\maketitle

\label{Sec:Introduction}
\textit{Introduction}---Microwave quantum communication is an important aspect of future quantum networks due to its compatibility with superconducting quantum circuits and modern telecommunication standards. Superconducting quantum circuits are currently one of the leading platforms in quantum information processing~\cite{Kjaergaard2020}, achieving a computational power beyond that of classical systems~\cite{Arute2019,Gao2025}. They operate at microwave frequencies around $\SI{5}{\giga\hertz}$ and in dilution refrigerators at temperatures around $\SI{10}{\milli\kelvin}$. While modern superconducting processors demonstrate coherent operations with more than one hundred superconducting qubits in individual dilution cryostats~\cite{Kim2023,Google2025}, it is well understood that millions of physical qubits will be required to tackle practical problems~\cite{Daley2022,Martinis2025}. However, the task of scaling up the number of superconducting qubits faces multiple challenges due to limitations in monolithic fabrication~\cite{Brecht2016}, spatial constraints~\cite{Hollister2024}, or available cooling power~\cite{Krinner2019}. A potential solution is to rely on distributed architectures~\cite{Bravyi2022} by connecting multiple cryogenic quantum nodes, with moderate numbers of qubits, into one large quantum network~\cite{Xiang2017,Penas2022}.

A particular approach to build such a quantum network is to couple remote quantum nodes directly at microwave carrier frequencies~\cite{Zhong2021,Abdo2025,Almanakly2025}. Recent successful realizations of deterministic microwave entanglement distribution between spatially-separated dilution refrigerators via cryolinks~\cite{Magnard2020,Yam2025} demonstrate the experimental feasibility and efficiency of this approach. Moreover, it has been demonstrated that microwave channels can be used for the distribution of quantum correlations at temperatures $T \gg \SI{10}{\milli\kelvin}$ as long as the channel intrinsic losses are sufficiently low~\cite{Yam2025,Qiu2025b}. These findings promise to relax technical requirements for future microwave quantum links and enable large-scale superconducting quantum networks. At the same time, noisy quantum networks are becoming a focus of intense research also at optical frequencies~\cite{Im2021,Coutinho2022,Grasselli2022} and for other material systems~\cite{Nickerson2014,Beckert2024,Lukin2025}.

In this Letter, we report on the experimental realization of a superconducting microwave quantum network with channel temperatures up to $\SI{4}{\kelvin}$, mimicking a ``hot" quantum channel compared to the millikelvin operation temperature of superconducting quantum circuits. We demonstrate its potential by distributing continuous-variable entanglement between two spatially-separated dilution refrigerators and exploiting it for coherent state teleportation~\cite{Bennett1993,Fedorov2021,Qiu2025a} with fidelities exceeding the classical threshold. Specifically, we use steady-state, two-mode squeezed (TMS) states as the entanglement resource to teleport coherent states over a $\SI{6.6}{\meter}$ long cryolink~\cite{Yam2025}, achieving teleported state fidelities up to $\SI{72.6 \pm 0.1}{\percent}$ at the channel base temperature of $\SI{170}{\milli\kelvin}$. We demonstrate that the quantum entanglement is preserved even when the TMS microwave states pass through a $\SI{4}{\kelvin}$ thermal channel and permits the deterministic quantum teleportation with fidelity of $\SI{59.9 \pm 2.5}{\percent}$. This key achievement of our work is enabled by the low microwave losses of the superconducting niobium-titanium (NbTi) cables~\cite{Kurpiers2017,Yam2025}, with critical temperature $T_\textrm{c} \simeq \SI{10}{\kelvin}$, used as the quantum channel. As a consequence, the fluctuation-dissipation theorem~\cite{Callen1951} dictates that propagating photonic modes are effectively decoupled from their ambient hot environment, thus preserving the quantum correlations therein.
\begin{figure*}[t]
    \includegraphics[width=\linewidth]{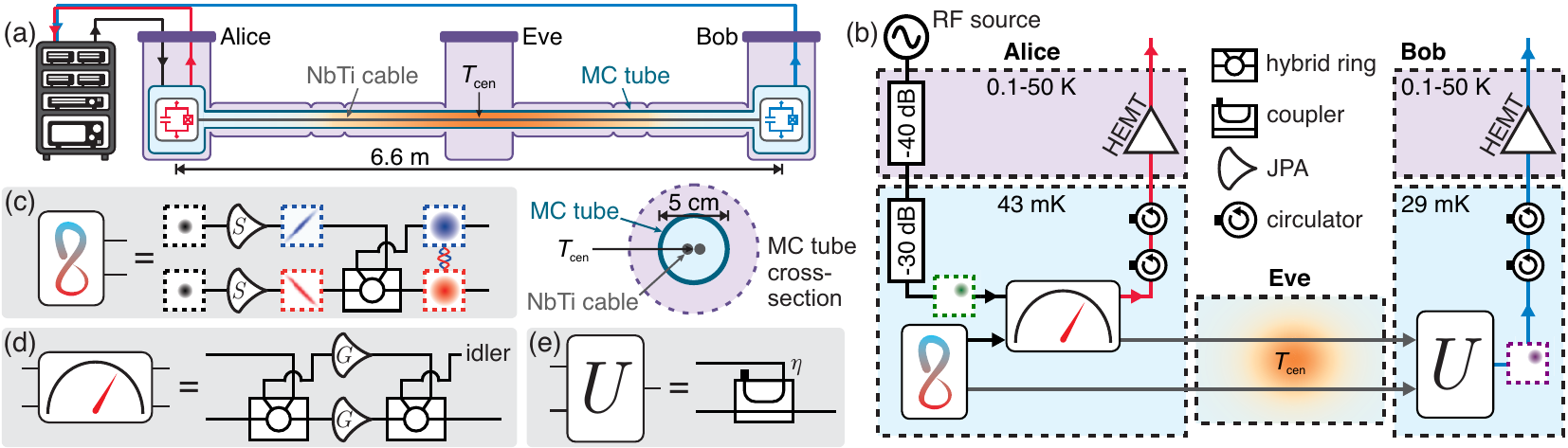}
    \caption{Cryolink and experimental setup. (a) Schematic of the cryolink connecting Alice and Bob, including a cross-sectional view of the MC tube at the cryolink center. (b) Schematic representation of the coherent quantum teleportation protocol with an analog feedforward. Schematics of (c) the TMS state generation, (d) the feedforward generation with a Josephson interferometer, and (e) Bob's local displacement operation.}
    \label{Fig1}
\end{figure*}

\label{Sec:ExperimentalSetup}
\textit{Experimental setup}---The cryolink used in our experiment consists of two dilution refrigerators (Alice and Bob) separated center-to-center by $\SI{6.6}{\meter}$ and connected via a cold network node (Eve)~\cite{Yam2025}. As illustrated in Fig.\,\ref{Fig1}(a), the quantum channel between the mixing chamber (MC) stages of Alice and Bob consists of 6-meter-long superconducting NbTi coaxial cables with characteristic losses of $\SI{1}{\decibel\per\kilo\meter}$. The channel temperature at its center, $T_\textrm{cen}$, can be varied in the range between the cryolink base temperature $\SI{170}{\milli\kelvin}$ and $\SI{4}{\kelvin}$ by using an embedded heater.

We generate and control microwave quantum states in Alice by using four flux-driven Josephson parametric amplifiers (JPAs)~\cite{Honasoge2025} in combination with linear microwave elements, such as hybrid rings, circulators, and directional couplers. The microwave signals propagate through two parallel superconducting coaxial cables towards Bob. These signals distribute quantum entanglement and provide the measurement feedforward, thereby implementing the quantum teleportation protocol~\cite{Fedorov2021}. In our experiment, we tune all four JPAs to the frequency $\omega_0/2\pi = \SI{5.35}{\giga\hertz}$ and perform squeezing operations by introducing a strong coherent pump tone at the frequency $\omega_\textrm{p} = 2\omega_0$~\cite{Zhong2013,Fedorov2018}. During state tomography, the output signals are amplified, down-converted, and detected with a field-programmable gate array (FPGA) receiver. The FPGA calculates signal statistical moments up to the fourth order in order to reconstruct the corresponding quantum states in the Gaussian approximation~\cite{Weedbrook2012} and verify their Gaussianity. We calibrate the gain and noise of our amplification chain using the Planck spectroscopy~\cite{Gandorfer2025}. More details about the devices and experimental setup are provided in the Supplemental Material~\cite{SupMat}.

\label{Sec:InterfridgeMicrowaveTeleportation}
\textit{Inter-fridge microwave teleportation}---Quantum teleportation is an entanglement-assisted communication protocol that allows disembodied transfer of unknown quantum states~\cite{Bennett1993}. While quantum teleportation of optical photonic states has been demonstrated in various experimental settings, including ground-to-satellite realizations~\cite{Ren2017}, quantum teleportation of microwave states has so far been constrained to experiments within individual cryogenic nodes~\cite{Fedorov2021,Qiu2025a}. Here, we successfully implement an analog (or measurement-device-free~\cite{Fedorov2021,Yamashima2025}) quantum teleportation of coherent microwave states between spatially-separated dilution refrigerators.

We employ two ``entanglement JPAs" [see Fig.\,\ref{Fig1}(c)] to generate a balanced TMS state, characterized by a squeezing level $S_\textrm{TMS}$, between Alice and Bob~\cite{Fedorov2018}. Then, we perform a Bell-type measurement operation on Alice's state, where the latter is obtained by superimposing an input coherent state and Alice's part of the TMS state via a hybrid ring (balanced microwave beam splitter). The measurement operation utilizes two ``measurement JPAs" [see Fig.\,\ref{Fig1}(d)] in a Josephson interferometer configuration~\cite{Kronowetter2023} in the regime of high degenerate gain $G \gg 1$~\cite{Renger2021}. The produced analog measurement result can be treated as a classical feedforward signal and is sent to Bob through a separate coaxial cable in the cryolink. Combined with Bob's part of the TMS state via a directional coupler with coupling $\eta$ [see Fig.\,\ref{Fig1}(e)], the feedforward signal determines the displacement amplitude and quadrature variance of a teleported state~\cite{Fedorov2016,Fedorov2021}. To produce a correct complex displacement amplitude, we require $G = 4\eta$ in the ideal case of zero losses in all experimental elements. Our particular directional coupler has $\eta = \SI{15}{\decibel}$, hence, we operate the measurement JPAs around $G = \SI{21}{\decibel}$. Bob's teleported  state at the directional coupler output, ideally, has the complex displacement perfectly matching that of Alice's input state and quadrature variances at the vacuum level.
\begin{figure}[t]
    \includegraphics[width=\linewidth]{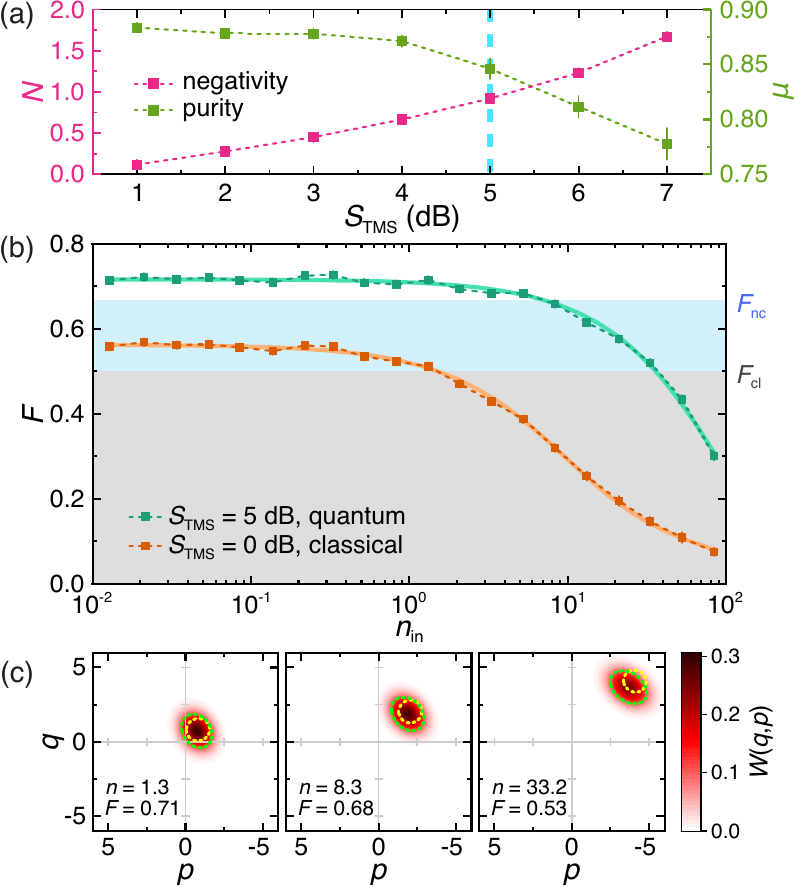}
    \caption{Quantum teleportation between distant dilution refrigerators. (a) Negativity $N$ and purity $\mu$ of TMS states distributed over the cryolink for various squeezing levels $S_\textrm{TMS}$. The vertical dashed line indicates the chosen operating point of $S_\textrm{TMS}=\SI{5}{\decibel}$. (b) Teleported state fidelities $F$ of ``quantum" and ``classical" teleportation protocols as a function of the input state photon number $n_\textrm{in}$ and averaged across displacement angles. Solid lines represent fits using our theoretical model. Gray- and blue-shaded regions indicate fidelities below the classical $F_\textrm{cl}$ and no-cloning $F_\textrm{nc}$ thresholds, respectively. (c) Exemplary Wigner functions of relevant coherent states. Yellow and green dashed lines denote the $1/e$ Gaussian contours of the input and teleported states, respectively. Error bars denote the standard error of the experimental data and are smaller than the symbol size when not shown.}
    \label{Fig2}
\end{figure}

Figure\,\ref{Fig2}(a) shows calibration measurements of our TMS resource states. Here, we sweep the squeezing level of the entanglement JPAs from \SI{1}{\decibel} to \SI{7}{\decibel} and observe that the negativity $N$ (purity $\mu$) increases (decreases) with increasing $S_\textrm{TMS}$. Negativity is an entanglement monotone derived from the PPT criterion, where $N>0$ implies entanglement~\cite{Peres1996,Adesso2005}, and purity describes how much classical noise is admixed into a quantum state, with $\mu = 1$ corresponding to a pure state. For high-fidelity quantum teleportation, we want both high $N$ and $\mu$. However, in our experiment, large microwave parametric pump powers (around $\SI{-40}{\dBm}$ at the JPA pump input) are needed to achieve squeezing levels around $\SI{5}{\decibel}$ below vacuum. Those strong pump signals introduce additional noise due to higher-order nonlinearities and photon number uncertainty in the JPAs~\cite{Boutin2017,Renger2021}. As a practical compromise, we use TMS states with $S_\textrm{TMS} = \SI{5}{\decibel}$, corresponding to $N = \SI{0.922 \pm 0.009}{}$ and $\mu = \SI{0.846 \pm 0.007}{}$.

The quantum teleportation protocol is typically characterized by its fidelity, $F$, which is defined as the overlap between Alice's input and Bob's output states~\cite{Weedbrook2012}. Figure\,\ref{Fig2}(b) shows the measured fidelities for teleported coherent states at the carrier frequency of $\SI{5.35}{\giga\hertz}$ with average photon numbers, $n_\textrm{in}$, in the range from $10^{-2}$ to $10^{2}$ and averaged across all displacement angles $\theta_\textrm{in} \in [0,2\pi)$. First, we consider the ``classical" teleportation case by turning off the entanglement JPAs, $S_\textrm{TMS} = \SI{0}{\decibel}$, to provide a fidelity reference for teleported states. We measure fidelities up to $\SI{56}{\percent}$ for coherent states with low photon numbers $n_\textrm{in} \ll 1$. For larger values of $n_\textrm{in} > 1$, the fidelities drop below the asymptotic classical threshold $F_\textrm{cl} = 1/2$~\cite{Braunstein2001}. Near-vacuum input states can be teleported with fidelities above $F_\textrm{cl}$ due to losses in the experimental setup, which admix vacuum into the teleported states.

For the teleportation protocol with finite quantum entanglement, $S_\textrm{TMS} = \SI{5}{\decibel}$, we observe fidelities exceeding those of the ``classical" case, thus, demonstrating a quantum advantage across the entire parameter range. We are able to teleport low photon number coherent states with fidelities of around $\SI{72}{\percent}$, reaching a maximum of $F = \SI{72.6 \pm 0.1}{\percent}$. This value substantially exceeds the asymptotic no-cloning fidelity of $F_\textrm{nc} = 2/3$~\cite{Grosshans2001}, which implies the possibility to guarantee unconditional security when using quantum teleportation as a communication protocol for a codebook of coherent states with $n_\textrm{in} \in [0,\infty)$. However, experimental imperfections, such as losses and JPA compression effects, cause $F$ to decrease with $n_\textrm{in}$, such that fidelities exceed $F_\textrm{nc}$ ($F_\textrm{cl}$) only up to $n_\textrm{in} = \SI{8.3}{}$ ($n_\textrm{in} = \SI{33.2}{}$). To understand this drop in $F$, we consider the reconstructed Wigner functions in Fig.\,\ref{Fig2}(c). From $n_\textrm{in} = \SI{1.3}{}$ to $n_\textrm{in} = \SI{33.2}{}$, the displacement mismatch between the teleported and input state distributions grows up to $\SI{0.554}{}$ photons, which corresponds to an estimated $\SI{15.9}{\percent}$ decrease in fidelity. Thus, the drop in $F$ is mainly caused by a mismatch in teleported state displacement, which results from losses in the experimental setup. This behavior is well-described by our theoretical model [solid lines in Fig.\,\ref{Fig2}(b)], which incorporates losses and noise to simulate the teleportation protocol~\cite{Holevo2007,Fedorov2021}. An analytic expression can be obtained for the teleported state fidelity
\begin{equation}\label{Eq:model}
    F(\alpha,\kappa,\zeta) = \frac{2}{\zeta + \kappa + 1} \exp\left[ -2\frac{(\sqrt{\kappa}-1)^2}{\zeta + \kappa + 1} |\alpha|^2 \right],
\end{equation}
\begin{equation}
    \zeta = (1+\kappa) \cosh 2r - 2\sqrt{\kappa} \sinh 2r + n_\textrm{dev} + n_\textrm{th},
\end{equation}
where $\alpha$ is the input coherent state amplitude, $\kappa$ the effective gain, $\zeta$ the effective added noise, $r$ the resource squeezing parameter, $n_\textrm{dev}$ the added device noise, and $n_\textrm{th}$ the coupled thermal noise. Fitting the teleportation data for $S_\textrm{TMS} = \SI{5}{\decibel}$ with this expression, we obtain $\kappa = 0.778$, which implies an overall attenuation of about $\SI{1.09}{\decibel}$ in our implementation of quantum teleportation, and $\zeta = 1.015$, which can be attributed to finite squeezing, noisy devices, and losses in the setup. In the current experiment, the largest source of infidelity for the teleported states, especially at large $n_\textrm{in}$, is the microwave insertion losses in passive microwave devices, such as the hybrid rings, circulators, and directional couplers. Substituting those with low-loss superconducting devices~\cite{Kannan2023,Navarathna2023} could substantially increase the teleportation fidelities by $\sim \SI{15}{\percent}$. More details about our model of the quantum teleportation protocol are provided in the Supplemental Material~\cite{SupMat}.

\begin{figure}[t]
    \includegraphics[width=\linewidth]{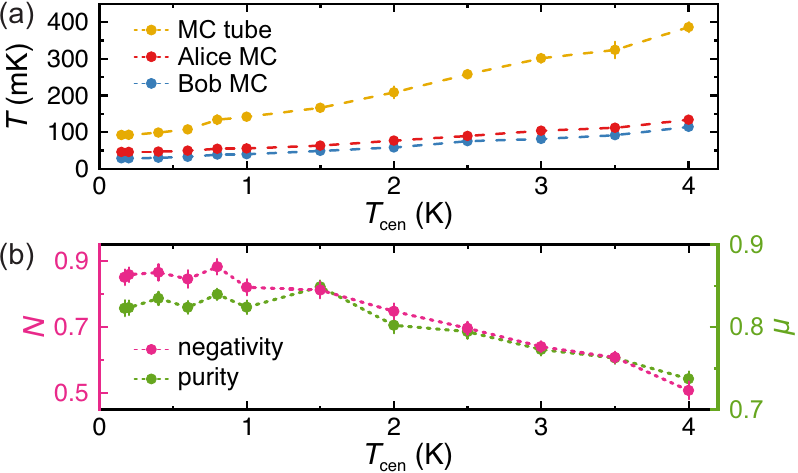}
    \caption{Entanglement distribution through the \SI{4}{\kelvin} thermal channel. (a) Average temperatures $T$ of various MC sections as a function of the cryolink center temperature $T_\textrm{cen}$. (b) Negativity $N$ and purity $\mu$ of the TMS states as a function of $T_\textrm{cen}$. Error bars denote the standard error of the experimental data and are smaller than the symbol size when not shown.}
    \label{Fig3}
\end{figure}

\label{Sec:ThermalTeleportation}
\textit{Teleportation through a thermal channel}---In the next step, we use the cryolink to mimic a noisy thermal communication channel by heating its center temperature up to $T_\textrm{cen} = \SI{4}{\kelvin}$. This is accomplished by using a PID-controlled heater clamped to the NbTi cables at the center of the cryolink. However, the local heating also affects the MC temperatures of the Alice and Bob cryostats, as shown in Fig.\,\ref{Fig3}(a), due to a direct (parasitic) thermal coupling to the heater. This thermal coupling eventually limits the teleportation fidelities at high channel temperatures. More details about the heating procedure and calibration are available in the Supplementary Material~\cite{SupMat}.

Figure\,\ref{Fig3}(b) shows the measured negativity and purity values of the TMS states distributed between Alice and Bob at various $T_\textrm{cen}$. We see that as $T_\textrm{cen}$ rises from $\SI{0.17}{\kelvin}$ to $\SI{4}{\kelvin}$, $N$ decreases from $\SI{0.851 \pm 0.024}{}$ to $\SI{0.508 \pm 0.027}{}$ and $\mu$ decreases from $\SI{0.823 \pm 0.010}{}$ to $\SI{0.737 \pm 0.009}{}$. This decrease in entanglement can be attributed to the additional thermal noise. Importantly, $N$ remains above zero, demonstrating that quantum entanglement survives even when passing through a hot quantum link.
\begin{figure}[t]
    \includegraphics[width=\linewidth]{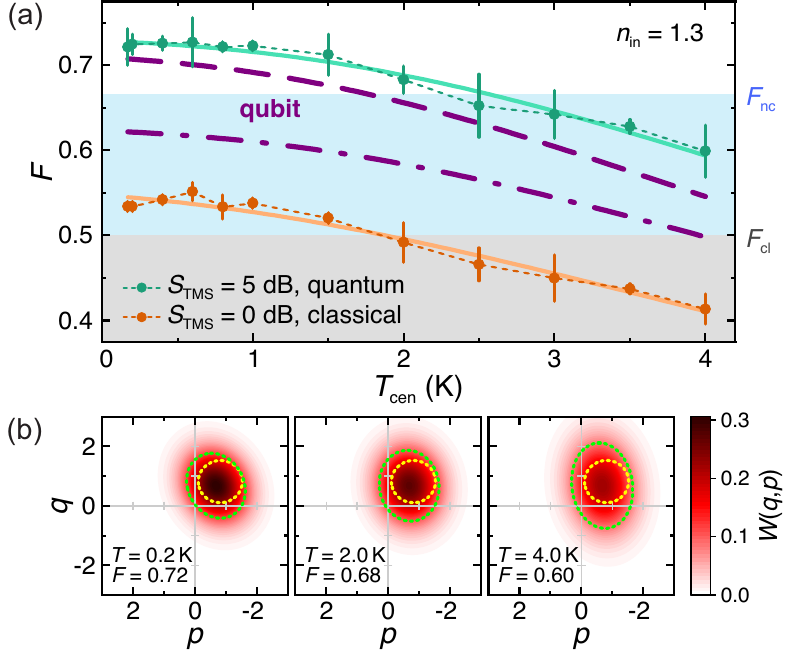}
    \caption{Quantum teleportation through the \SI{4}{\kelvin} thermal channel. (a) Teleported state fidelities $F$ of ``quantum" and ``classical" teleportation for various $T_\textrm{cen}$. Solid lines represent fits using our theoretical model. Gray- and blue-shaded regions indicate fidelities below the classical $F_\textrm{cl}$ and no-cloning $F_\textrm{nc}$ thresholds, respectively. Purple dash-dotted and dashed lines represent predicted qubit state teleportation fidelities for $S_\textrm{TMS}=\SI{5}{\decibel}$ and $S_\textrm{TMS}=\SI{10}{\decibel}$, respectively. (b) Wigner functions of exemplary teleported coherent states. Yellow and green dashed lines denote the $1/e$ contours of the input and teleported states, respectively. Error bars denote the standard error of the experimental data and are smaller than the symbol size when not shown.}
    \label{Fig4}
\end{figure}

Next, we perform microwave quantum teleportation over the described thermal channel. Figure\,\ref{Fig4}(a) shows the teleported coherent state fidelities $F$ for $n_\textrm{in} = 1.3$ as a function of $T_\textrm{cen}$. We observe that the teleported state fidelities stay above the no-cloning limit $F > F_\textrm{nc}$ up to $T_\textrm{cen} \sim \SI{2}{\kelvin}$ and then decrease to $F = \SI{59.9 \pm 2.5}{\percent}$ at $T_\textrm{cen} = \SI{4}{\kelvin}$. In the whole temperature range, the quantum teleportation fidelities consistently remain above the ``classical" teleportation fidelities, demonstrating an overall quantum advantage of the implemented communication protocol. We fit the collected data using Eq.\,(\ref{Eq:model}), taking into account the increased Alice MC temperatures and the resulting thermal noise $n_\textrm{th}$. The outcome is shown by solid lines in Fig.\,\ref{Fig4}(a). Observing the reconstructed Wigner functions in Fig.\,\ref{Fig4}(b), we see that when sweeping $T_\textrm{cen} = \SI{0.2}{\kelvin}$ to $T_\textrm{cen} = \SI{4}{\kelvin}$, the teleported state variance increases up to $\SI{0.412}{}$ photons, which implies an estimated $\SI{13.6}{\percent}$ decrease in $F$. Within this variance change, the lossy components in the Alice (quantum link) setup contribute $\SI{0.38}{}$ ($\SI{0.02}{}$) of added noise photons, and hence, a $\SI{12.7}{\percent}$ ($\SI{0.8}{\percent}$) decrease in $F$. Thus, we find that microwave quantum communication over the cryolink is robust against its center temperature. This observation can be explained by the fluctuation-dissipation theorem~\cite{Callen1951} and interpreted using the beam splitter model from quantum optics~\cite{Weedbrook2012}. Here, the thermal noise coupled to a quantum channel is described by its mean photon number, $n_\textrm{th} = \varepsilon n_\textrm{env}$, where $\varepsilon$ is the channel losses and $n_\textrm{env}$ is the mean noise photon number of the ambient thermal environment. For vanishing losses $\varepsilon \to 0$, the coupled noise $n_\textrm{th}$ is strongly suppressed, despite a hot environment with $n_\textrm{env} \gg 1$. The current limitation of the demonstrated protocol rather stems from the parasitic temperature increase of the Alice MC itself. For better teleportation fidelities at high $T_\textrm{cen}$, one has to improve thermal isolation between the cold nodes (Alice and Bob) and the hot cryolink center by exploiting various cryogenic techniques, such as heat switches~\cite{Shu2022}, on-demand thermal couplers~\cite{Qiu2025b}, multilayer insulation against black-body radiation, and thermal breaks.

\label{Sec:QubitTeleportation}
\textit{Qubit state teleportation}---Finally, we theoretically examine implications of our continuous-variable protocol for a hybrid quantum teleportation of an input qubit state $\ket{\psi} = \cos\frac{\theta}{2} \ket{0} + e^{i\varphi} \sin\frac{\theta}{2}\ket{1}$. Since the TMS state contains contributions from all photon number states, it is theoretically possible to employ it for teleportation of qubits. We analyze the performance of this hybrid teleportation protocol using fidelity $F(\theta,\kappa,\zeta)$ in terms of the aforementioned parameters $\kappa$ and $\zeta$. We follow the formalism provided in Refs.\,\cite{Takeda2013b, Renger2023} to find expressions for the ground state $\ket{0}$ teleportation fidelity
\begin{equation}
	F(0,\kappa,\zeta) = \frac{2}{\zeta+\kappa+1},
\end{equation}
which recovers the scenario of teleporting a coherent state with $\alpha = 0$, and the excited state $\ket{1}$ teleportation fidelity
\begin{equation}
	F(\pi,\kappa,\zeta) =  \frac{2\zeta^2 - 2\kappa^2 + 12\kappa - 2}{(\zeta+\kappa+1)^3}.
\end{equation}
In order to quantify the hybrid teleportation of an arbitrary pure qubit state, we integrate over $\theta$ and $\varphi$ to find the average qubit state teleportation fidelity
\begin{equation}
    \begin{split}
    	\bar{F}(\kappa,\zeta) &= \frac{1}{4\pi} \int_{0}^{2\pi} \int_{0}^{\pi} F(\theta,\kappa,\zeta) \sin\theta ~d\theta d\varphi \\
        &= \frac{6\zeta + 4\sqrt{\kappa}}{3(\zeta+\kappa+1)^2} + \frac{16\kappa}{3(\zeta+\kappa+1)^3}.
    \end{split}
\end{equation}
Now, we can use the parameters obtained from fitting the experimental data of coherent state teleportation to predict $\bar{F}(\kappa,\zeta)$ using our setup, as shown by the purple dash-dotted and dashed lines in Fig.\,\ref{Fig4}(a). We estimate the average qubit teleportation fidelities of $\bar{F} = \SI{62.2}{\percent}$ for $S_\textrm{TMS}=\SI{5}{\decibel}$, and $\bar{F} = \SI{70.7}{\percent}$ when $S_\textrm{TMS}=\SI{10}{\decibel}$, which are experimentally feasible squeezing levels with our JPAs~\cite{Honasoge2025}. Although these fidelities are not yet as high as they have to be for direct applications in distributed quantum computing, they nevertheless represent the potential and universality of hybrid quantum communication protocols that combine continuous-variable and discrete-variable states. More details about our model for the hybrid quantum teleportation and its potential improvements are available in the Supplemental Material~\cite{SupMat}.

\label{Sec:Discussion}
\textit{Discussion}---Our realization of quantum teleportation over a noisy quantum channel demonstrates the potential of microwave communication in practical quantum networks. We successfully teleport microwave states across spatially-separated dilution refrigerators, achieving fidelities of $F > F_\textrm{nc}$ for $n_\textrm{in} \le \SI{8.3}{}$ in the low temperature regime, $T \lesssim \SI{1}{\kelvin}$. At the same time, in the high temperature regime, teleportation fidelities degrade due to a parasitic coupling to the channel heater but still remain above the classical threshold, $F > F_\textrm{cl}$, even at $T_\textrm{cen} = \SI{4}{\kelvin}$. The capabilities to extend beyond a single dilution refrigerator and operate at high temperatures are crucial for incorporating superconducting circuits into heterogeneous quantum networks~\cite{Clerk2020}. Furthermore, one can enhance the demonstrated experimental techniques by combining the Gaussian entanglement resources with non-Gaussian operations provided by qubits, such as teleporting Fock states~\cite{Takeda2013a}, entanglement distillation of TMS states~\cite{Kurochkin2014}, or autonomous qubit entanglement~\cite{Agusti2023}. In a more general context, our results demonstrate that thermal equilibrium fluctuations do not necessarily limit the distribution of quantum entanglement through noisy quantum channels, as long as the channel losses are low. It also motivates further investigation of quantum networks in non-equilibrium scenarios, related to the question about how quantum states thermalize in distributed quantum systems.

\textit{Acknowledgments}---We acknowledge support by the German Research Foundation via Germany`s Excellence Strategy (EXC-2111-390814868), the EU Quantum Flagship project QMiCS (Grant No.~820505), the German Federal Ministry of Education and Research via the project QUARATE (Grant No.~13N15380). This research is part of the Munich Quantum Valley, which is supported by the Bavarian state government with funds from the Hightech Agenda Bayern Plus.

\bibliography{Bibliography} 

\end{document}